\documentclass[12pt]{article}

\usepackage{amsmath,amsfonts,latexsym,amssymb,amscd}
\usepackage{pslatex}
\usepackage[latin1]{inputenc}
\usepackage[T1]{fontenc}
\usepackage{pspicture}
\usepackage{verbatim,amsthm,curves,graphics}
\usepackage{mathrsfs}

\usepackage{graphicx}

\newcommand{\G}{{\mathcal G}}
\newcommand{\mr}{{\mathbb R}}

\newcommand{\mc}{{\mathbb C}}
\newcommand{\mz}{{\mathbb Z}}

\begin{document}


\title{Uniqueness and nonuniqueness  of the stationary black holes  in 5D Einstein-Maxwell and Einstein-Maxwell-dilaton gravity }

\author{
Stoytcho Yazadjiev$^{}$\thanks{\tt yazad@phys.uni-sofia.bg}
\\ \\
{\it $ $Department of Theoretical Physics, Faculty of Physics, Sofia
University} \\
{\it 5 J. Bourchier Blvd., Sofia 1164, Bulgaria} \\
    }
\date{}

\maketitle

\begin{abstract}
In the present paper we investigate the general problem of uniqueness of the stationary black solutions in 5D Einstein-Maxwell-dilaton gravity with arbitrary dilaton coupling parameter
containing the Einstein-Maxwell gravity as a particular case.  We formulate and prove uniqueness theorems classifying the stationary black hole solutions in terms of their
interval structure, electric and magnetic charges and the magnetic fluxes. The proofs are based on  the nonpositivity of the Riemann curvature operator on the space of the potentials
which imposes restrictions on the dilaton coupling parameter.
\end{abstract}


\sloppy

\section{Introduction}

The stationary, asymptotically flat  black hole solutions in 5-dimensional Einstein gravity with two axial Killing fields were classified
in \cite{HY1}. The vacuum black hole solutions were classified in terms of their interval structure and  angular
momenta. The generalization of \cite{HY1} for a certain sector of 5D Einstein-
Maxwell gravity was done in \cite{HY2}.  In the sector under consideration, the 5D asymptotically
flat Einstein-Maxwell black holes are classified in terms of their interval structure, angular
momentum and the magnetic charges associated with the generators of $H_2(M)$. Straightforward extensions of
\cite{HY1} and \cite{HY2} to the case of asymptotically flat black objects in 5D minimal supergravity were given in
\cite{TYI1}--\cite{AH}.

However, the problem of classification of
the asymptotically flat stationary Einstein-Maxwell black holes in the general case is still open --  it is not known whether
the   Einstein-Maxwell black holes are uniquely specified in terms of their interval structure, angular momentum, electric and magnetic charges.
In the present paper we address this question in the more general context of Einstein-Maxwell-dilaton gravity with an arbitrary dilaton coupling parameter $\alpha$.
The Einstein-Maxwell gravity is recovered by first putting $\alpha=0$ and then $\varphi=0$.

In the proofs of the classification theorems of
\cite{HY1}--\cite{AH} we strongly rely on the higher degree of symmetries of the dimensionally reduced stationary and axisymmetric field equations.
In the cases of  \cite{HY1} -- \cite{AH} the space of potentials is a symmetric space which insures the existence of nice
properties and, in particular, insures the existence of the so-called Mazur identity \cite{Mazur}, which is a key point in the  proofs.
The higher degree of symmetry of the dimensionally reduced  field equations is  a luxury that is not  expected to occur
in the general case. Even in 4 dimensions the space of potentials of the dimensionally reduced Einstein-Maxwell-dilaton gravity is not
a symmetric space in the general case \cite{Yazadjiev}. As we will see below this also is the case of 5D Einstein-Maxwell gravity and Einstein-Maxwell-dilaton gravity
for $\alpha^2\ne 8/3$. This fact makes 5D Einstein-Maxwell-dilaton gravity (and 5D Einstein-Maxwell gravity ) much more difficult to be investigated than
the 5D minimal supergravity which potential space is highly symmetric\footnote{This fact makes the finding of exact solutions in  5D Einstein-Maxwell and 5D Einstein-Maxwell-dilaton gravity with arbitrary $\alpha$ very difficult.
Up to now even the solutions describing rotating black holes with spherical horizon topology are not found contrary to the 5D minimal supegravity where exact
solutions are easy to be generated.} (see for example \cite{BCCGSW}). Therefore the overall strategy of \cite{HY1} and  \cite{HY2} is not applicable to the case under
consideration in the present paper.
That is why we use here the strategy of  \cite{Yazadjiev} which turned out successful in four dimensions.

It is worth mentioning that the difficulties are hidden not only in the mathematical tools for proving the uniqueness theorems but they are also hidden in the very
formulation of the uniqueness theorems.  It turns out that  the interval structure and the charges are insufficient to uniquely determine the
black hole solutions  in the general case.  A priori it is not clear what kind of  further suitable parameters associated with
the black solutions should be specified in addition to the interval structure and charges. In the present paper we show that, in the general case,
the suitable additional parameters are the magnetic fluxes. The role of the magnetic fluxes in the uniqueness theorems and the thermodynamics was first
noticed in \cite{YN1}, \cite{YN2} and \cite{Y}. In fact we borrow the construction of \cite{YN1}-- \cite{Y} in order to introduce the
magnetic fluxes in the present context.

The paper is organized as follows. In Sec. I   we give in concise form the necessary mathematical
base. In Sec. II  we present the dimensionally reduced  Einstein-Maxwell-dilaton equations and their formulation in terms of harmonic maps.
The main results are presented in Sec. III and Sec. IV. The paper ends with conclusion where we give  the natural extension of our results
for spacetimes with multiple  disconnected  horizons and comment on some other possible extensions.

\section{Stationary Einstein-Maxwell-dilaton black holes in 5D}

Let $(M,g_{ab},F_{ab})$ be  $5$-dimensional, analytic, asymptotically flat,
stationary black hole spacetime satisfying the Einstein-Maxwell-dilaton
equations
\begin{eqnarray}\label{FE}
&&R_{ab}= 2\partial_a\varphi \partial_b\varphi +    2 e^{-2\alpha\varphi}\left(F_{ac}F_{b}{}^c- \frac{g_{ab}}{6}F_{cd}F^{cd} \right),\\
&&\nabla_{a}\left( e^{-2\alpha\varphi }F^{ab}\right)=0=\nabla_{[a} F_{bc]}, \\
&&\nabla_{a}\nabla^a \varphi = - \frac{\alpha}{2} e^{-2\alpha\varphi} F_{cd}F^{cd} .
\end{eqnarray}
Let $\xi$ be the
asymptotically timelike complete Killing field, $\pounds_\xi g = 0$,
which we assume is normalized  near infinity,  $\lim_{\infty} \,  g(\xi,\xi) = -1$. We assume also that
the Maxwell and the dilaton  fields are  invariant under $\xi$, i.e.  $\pounds_\xi F = 0$ and $\pounds_\xi \varphi=0$.
We denote by
$H = \partial B$ the horizon of the black hole, where the black hole $B$
is defined as usual by $B = M \setminus
I^-({\mathcal J}^+)$, with ${\mathcal J}^\pm$ being the  null-infinities of
the spacetime.
We assume that $H$ is ``non-degenerate''
and that the horizon cross section is a compact connected
manifold of dimension $3$. We also assume that  the black hole exterior, $\langle\langle M\rangle\rangle$, is globally hyperbolic. By the topological
censorship theorem \cite{Galloway} the exterior $\langle\langle M\rangle\rangle$ is simply connected manifold with  boundary $\partial \langle\langle M\rangle\rangle=H$.

Here we will assume the existence of 2 further axial Killing fields $\eta_1$ and $\eta_2$ which are
mutually commuting and commute with the asymptotically timelike Killing field $\xi$, have
periodic orbits with period $2\pi$ and leave the Maxwell and dilaton fleld invariant, i.e. $\pounds_{\eta_1} F=\pounds_{\eta_2} F= 0$ and
$\pounds_{\eta_1} \varphi=\pounds_{\eta_2} \varphi= 0$ . The group of isometries is hence ${\cal G}=\mr\times U(1)^2$, where $\mr$ stands for the flow of $\xi$ while
$U(1)^2$ corresponds to the commuting flows of axial Killing fields.

Due to the symmetries of the spacetime the natural space to work on is the orbit (factor)
space ${\hat M}=\langle\langle M\rangle\rangle/{\cal G}$, where ${\cal G}$ is the isometry group. The structure of the factor
space is described by the following theorem ( see \cite{HY1} and \cite{HY3}):

\medskip
\noindent
{\bf Theorem:} Let $(M,g)$ be a stationary, asymptotically Einstein-Maxwell-dilaton, 5-dimensional black hole spacetime with isometry group
$\G=\mr\times U(1)^2$ satisfying the technical assumptions stated above. Then the orbit space ${\hat M}=\langle\langle M\rangle\rangle /{\cal G}$ is
a 2-dimensional manifold with boundaries and corners homeomorphic to a half-plane. One boundary segment $I_{H}\subset \partial {\hat M}$
corresponds to the quotient of the horizon ${\cal H}=H/\G$, while the remaining segments $I_j$ correspond to the various axes, where
a linear combination $a_1(I_j) \eta_1 + a_2 (I_j)\eta_2=0$ and  ${\bf a}(I_j)=(a_1(I_j),a_2(I_j))\in \mz^2$.  For adjacent intervals $I_j$
and $I_{j+1}$ (not including the horizon) the vectors ${\bf a}(I)=(a_1(I),a_2(I))$  are subject to the following constraint
\begin{eqnarray}\label{constraint}
|\det \left(
       \begin{array}{cc}
         a_1(I_j) &  a_1(I_{j+1})\\
          a_2(I_j)& a_2(I_{j+1}) \\
       \end{array}
     \right)|=1 .
\end{eqnarray}
\medskip
\noindent

In the interior of ${\hat M}$ there is a naturally induced metric ${\hat g}$ which has signature $++$. We denote derivative operator associated with
${\hat g}$ by ${\hat D}$. Let us now consider the Gramm matrix of the Killing fields $G_{IJ}=g(K_{I},K_{J})$,
where $K_{0}=\xi$, $K_{1}=\eta_1$ and $K_{2}=\eta_2$. Then the determinant $\rho^2=|\det G|$ defines a scalar
function $\rho$ on ${\hat M}$ which, as  well known, is harmonic,
${\hat D}^{a}{\hat D}_{a}\rho=0$ as a consequence of the Einstein-Maxwell-dilaton field equations.
It can be shown that $\rho>0$, ${\hat D}_a \rho\ne 0$ in the interior of ${\hat M}$ and that $\rho=0$ on $\partial {\hat M}$.
We may define a conjugate harmonic function $z$ on ${\hat M}$ by $dz={\hat \star}\, d\rho$, where ${\hat \star}$ is the Hodge dual on ${\hat M}$.
The functions $\rho$ and $z$ define global coordinates on ${\hat M}$
identifying the orbit space with the upper complex half-plane
\begin{eqnarray}
{\hat M}= \{z+ i\rho \in \mc, \rho\ge 0  \}
\end{eqnarray}
with the boundary corresponding to the real axis. The induced metric ${\hat g}$ is given in these coordinates by

\begin{eqnarray}
{\hat g}= \Omega^2(\rho,z)(d\rho^2 + dz^2),
\end{eqnarray}
$\Omega^2$ being a conformal factor.

The above theorem allows us to introduce the notion of {\it interval structure}. The orbit space of the domain of outer communication
by the isometry group is a half plane ${\hat M}=\{z+i\rho, \rho>0\}$ and its boundary $\partial {\hat M}$ is divided into a finite number of
intervals $I_j$:

\begin{eqnarray}\label{zintervals}
(-\infty,z_1), (z_1,z_2),...,(z_N,z_{N+1}),(z_{N+1},+\infty) .
\end{eqnarray}

 To each interval we associate its length $l(I_j)$ and a vector ${\bf a}(I_j)=(a_1(I_j),a_2(I_j))\in \mz^2$ (subject to (\ref{constraint}))
when the interval does not correspond to a horizon. To the interval corresponding to the orbit space ${\cal H}$ of the horizon we associate
zero vector $(0,0)$. \emph{The data ${l(I_j)}$ together with  ${\bf a}(I_j)=(a_1(I_j), a_2(I_j))$  are called interval structure.} The vectors
${\bf a}(I_j)= (a_1(I_j),a_2(I_j))$ corresponding to the outermost intervals $(-\infty,z_1)$ and $(z_{N+1},+\infty)$ must be $(1,0)$  and $(0,1)$
 since the spacetime is asymptotically flat.

Furthermore, we have the following theorem about the topology of the horizons \cite{HY1,HY3}:

\medskip
\noindent

{\bf Theorem:} Under the assumptions made above the horizon cross section $H$ must be topologically
either $S^2\times S^1$, $S^3$ or a Lens space $L(p,q)$ ($p,q \in \mz$). Here $p$ is given by  $p=\det({\bf a}_{h-1}, {\bf a}_{h+1})$
where ${\bf a}_{h-1}$ and ${\bf a}_{h+1}$ are vectors adjacent to the  horizon ${\cal H}$. The topology of $H$
is $S^2\times S^1$ for $p=0$, $S^3$ for $p=\pm 1$ and $L(p,q)$ in the other cases.
\medskip
\noindent

\section{Dimensionally reduced  Einstein-Maxwel-dilaton \\equations}

We begin with the dimensional reduction of the Maxwell 2-form $F$. The invariance of $F$ under the flow of the Killing fields $\eta_I$ ($I=1,2$) gives
$0=\pounds_{\eta_I}F=di_{\eta_I} F + i_{\eta_I}dF=di_{\eta_I} F$. Since $\langle\langle M\rangle\rangle$ is simply connected there exist globally defined potentials
$\Phi_I$ such that $i_{\eta_I}F=d\Phi_I$.    Defining the matrix field $H_{IJ}$ by

\begin{eqnarray}
H_{IJ}= g(\eta_I,\eta_J)=H_{JI}
\end{eqnarray}
the Maxwell 2-form can be written as follows

\begin{eqnarray}
F= H^{IJ} \eta_I \wedge d\Phi_J + F_{\perp}
\end{eqnarray}
where $H^{IJ}$ is the inverse of $H_{IJ}$ and $F_{\perp}$ is a 2-form orthogonal of $\eta_1$ and $\eta_2$, i.e. $i_{\eta_{I}}F_{\perp}=0$.
From the very definition of the  2-form $F_{\perp}$ it follows that it has the form $F_{\perp}= \star \left(B\wedge \eta_1\wedge \eta_2 \right)$
where $B$ is an 1-form. Hence we find $-i_{\eta_2}i_{\eta_1}\star F= h B$ where
\begin{eqnarray}
h=\det(H_{IJ}).
\end{eqnarray}
Now we consider the 1-form $e^{-2\alpha\varphi}hB$
and take the exterior derivative of it, $d\left(e^{-2\alpha\varphi}hB\right)= - di_{\eta_2}i_{\eta_1}\left(e^{-2\alpha\varphi }\star F\right)$. Taking into account that
$\eta_1$ and $\eta_2$ commute we obtain $d\left(e^{-2\alpha\varphi}hB\right)= - i_{\eta_2}i_{\eta_1}d\left(e^{-2\alpha\varphi }\star F\right)$  which, in view of the field
equations (\ref{FE}), means that $d\left(e^{-2\alpha\varphi}hB\right)=0$. Using again the fact that $\langle\langle M\rangle\rangle$ is simply connected we conclude
that there exists a globally defined potential $\Psi$ such that $B=e^{2\alpha\varphi} h^{-1} d\Psi $. Therefore the Maxell  2-form  $F$ is completely determined by the potentials
$\Phi_I$ and $\Psi$, i.e.

\begin{eqnarray}
F= H^{IJ} \eta_I \wedge d\Phi_J + h^{-1} e^{2\alpha\varphi}\star \left(d\Psi \wedge \eta_1\wedge \eta_2\right) .
\end{eqnarray}
It is easy to see that the potentials $\Phi_I$ and $\Psi$ are invariant under the spacetime symmetries and therefore are naturally defined on the orbit space ${\hat M}$.

Our next step is to consider the twist 1-forms $\omega_I$ defined by

\begin{eqnarray}
\omega_I= \star\left(\eta_1 \wedge \eta_2 \wedge d\eta_I \right)=i_{\eta_2}i_{\eta_1}\star d\eta_I.
\end{eqnarray}
Since $\eta_1$ and $\eta_2$ commute  we find

\begin{eqnarray}
d\omega_I= i_{\eta_2}i_{\eta_1}\star \left(d^{\dagger}d\eta_I\right)=2i_{\eta_2}i_{\eta_1}\star R[\eta_I].
\end{eqnarray}
Making use of the field equations (\ref{FE}) we obtain

\begin{eqnarray}
d\omega_I = 4 i_{\eta_I}F\wedge i_{\eta_2}i_{\eta_1}\left(e^{-2\alpha\varphi} \star F\right)= 4d\Phi_I\wedge d\Psi .
\end{eqnarray}
Hence we conclude that there are globally defined potentials $\chi_I$ such that

\begin{eqnarray}\label{omega}
\omega_I = d\chi_I + 2\Phi_I d\Psi - 2 \Psi d\Phi_I.
\end{eqnarray}

The potentials $\chi_I$ are invariant under the flow of the Killing fields and therefore they are defined on the orbit space. In what follows it turns out
convenient to introduce also another potential $\sigma_I$ defined by

\begin{eqnarray}
\sigma_I= \chi_I - 2\Psi\Phi_I
\end{eqnarray}
and satisfying

\begin{eqnarray}\label{dsigma}
d\sigma_I= \omega_I - 4\Phi_Id\Psi .
\end{eqnarray}

The Einstein-Maxwell-dilaton equations in spacetime  with $\mr \times U(1)^2$ isometry group  are equivalent to
two decoupled groups of partial differential equations on the orbit space ${\hat M}$. The first group of nonlinear partial differential equations
is for the matrix field $H_{IJ}$ and the  potentials $\Phi_{I}, \Psi, \chi_{I}$:

\begin{eqnarray}
&&\rho^{-1}{\hat  D}_a\left(\rho {\hat D}^a H_{IJ}\right) = H^{KL}{\hat D}_a H_{IK} {\hat D}^a H_{JL} - h^{-1}\omega_{I}^{a}{\omega_{J}}_{a}
- 4 e^{-2\alpha\varphi}{\hat D}_{a}\Phi_{I}{\hat D}^{a}\Phi_{J}   \nonumber \\ &&  + \frac{4}{3}H_{IJ}\left(e^{-2\alpha\varphi}H^{KL}{\hat D}_{a}\Phi_{K}{\hat D}^{a}\Phi_{L} -
h^{-1} e^{2\alpha\varphi}{\hat D}_a\Psi {\hat D}^a\Psi  \right) \label{RFE1} , \\ \nonumber \\
&&\rho^{-1}{\hat D}_a \left(\rho\omega^{a}_{I}\right)= H^{KL}\omega^{a}_{K}{\hat D}_a H_{IL}  + h^{-1}\omega^{a}_{I}{\hat D}_a h, \\ \nonumber \\
&&\rho^{-1}{\hat D}_a\left( \rho e^{-2\alpha\varphi}{\hat D}^a\Phi_{I}\right)= e^{-2\alpha\varphi} H^{KL}{\hat D}_a\Phi_{K} {\hat D}^a H_{LI} -
h^{-1}\omega^{a}_I{\hat D}_{a}\Psi ,\\ \nonumber \\
&&\rho^{-1}{\hat D}_a\left( \rho e^{2\alpha\varphi}{\hat D}^a\Psi\right)= H^{KL}\omega^{a}_{K}{\hat D}_a\Phi_{L} + h^{-1} e^{2\alpha\varphi}{\hat D}_a\Psi {\hat D}^a h, \\
\nonumber \\
&&\rho^{-1}{\hat D}_a\left( \rho {\hat D}^a\varphi\right) = - \alpha \left(e^{-2\alpha \varphi }H^{KL} {\hat D}_a\Phi_{K} {\hat D}^a\Phi_{L} -
h^{-1}e^{2\alpha\varphi} {\hat D}_a\Psi {\hat D}^a\Psi \right),\label{RFE5}
\end{eqnarray}

Since  Eqs. (\ref{RFE1}) -- (\ref{RFE5}) are invariant under the conformal transformations of ${\hat g}_{ab}$ and since a 2-dimensional metric is conformally flat
they are decoupled from the second group equations for the function  $e^{2\Gamma}=h g(\nabla\rho,\nabla\rho)$ and  the covariant derivative $\hat D$
can be replaced by the covariant derivative $D$ for the flat metric $\delta_{ab}$ .

The second group  is for the function $e^{2\Gamma}=h g(\nabla\rho,\nabla\rho)$ on the orbit space ${\hat M}$ and written in terms of $D$ is given by:

\begin{eqnarray}
2\rho^{-1}D^{a}\rho D_a\Gamma= \left[2 D^a\varphi D^b\varphi + 2 e^{-2\alpha\varphi} H^{IJ}D^a\Phi_I D^b \Phi_J + 2e^{2\alpha\varphi} h^{-1} D^a\Psi D^b\Psi
\right.  \nonumber \\ \left.+ \frac{1}{2} h^{-1} H^{IJ}\omega_I^a\omega_J^b + \frac{1}{4}h^{-1} D^a h D^b h + \frac{1}{4} Tr \left(H^{-1}D^a H H^{-1} D^b H\right) \right]
\left[ \delta_{ab} - 2 D_a z D_b z\right],  \label{RFE6}\\   \nonumber \\
\rho^{-1}D^a z D_a\Gamma = \left[2 D^a\varphi D^b\varphi + 2 e^{-2\alpha\varphi} H^{IJ}D^a\Phi_I D^b \Phi_J + 2e^{2\alpha\varphi} h^{-1} D^a\Psi D^b\Psi
\right.  \nonumber \\ \left.+ \frac{1}{2} h^{-1} H^{IJ}\omega_I^a \omega_J^b + \frac{1}{4}h^{-1} D^a h D^b h + \frac{1}{4} Tr \left(H^{-1}D^a H H^{-1} D^b H\right) \right]
D_a \rho D_b z .\label{RFE7}
\end{eqnarray}

 Once the solution of the system
equations (\ref{RFE1}) - (\ref{RFE5}) is known we can  determine the function $\Gamma$.
 Therefore the problem of the classification  of the 5D Einstein-Maxwell-dilaton black hole solutions  can be studied as 2-dimensional boundary value problem for the nonlinear
 partially differential equation system   (\ref{RFE1}) -- (\ref{RFE5}) as the boundary conditions are specified below.

Now let us consider the 9-dimensional manifold ${\cal N}=\{(H_{IJ}(I\le J),\chi_{I},\Phi_{I},\Psi, \varphi) \in \mr^9; h>0 \} $ with the strictly positive definite
metric

\begin{eqnarray}
&&dL^2= G_{AB} \,dX^AdX^B= \frac{1}{4}Tr\left(H^{-1} dHH^{-1}dH\right) + \frac{1}{4}h^{-2}dh^2  \nonumber \\ && +
\frac{1}{2}h^{-1}H^{IJ}\left(d\chi_I + 2\Phi_I d\Psi - 2 \Psi d\Phi_I\right) \left(d\chi_J + 2\Phi_J d\Psi - 2 \Psi d\Phi_J\right)  \\ && + 2e^{-2\alpha\varphi} H^{IJ}d\Phi_{I}d\Phi_{J}
 + 2h^{-1}e^{2\alpha\varphi} d\Psi^2 +2 d\varphi^2  .\nonumber
\end{eqnarray}
If we use the potentials $\sigma_I$ instead of $\chi_I$ we have to replace $\omega_I= d\chi_I + 2\Phi_I d\Psi - 2 \Psi d\Phi_I$ by $\omega_I=d\sigma_I + 4\Phi_Id\Psi$
as follows from (\ref{dsigma}).

 Eqs. (\ref{RFE1}) -- (\ref{RFE5}) can be obtained from a variational principle based on the functional

\begin{eqnarray}\label{FUNCTIONAL}
I[X^A]= \int_{{\hat M}}d^2x\sqrt{-{\hat g}} {\hat g}^{ab}G_{AB}(X^C){\hat D}_aX^{A} {\hat D}_bX^{B}.
\end{eqnarray}

Further we consider the mapping
\begin{eqnarray}\label{harmonicmapp}
 {\cal X}: {\hat M} \mapsto {\cal N}
\end{eqnarray}
of the 2-dimensional Riemannian manifold  ${\hat  M}$ onto the 9-dimensional Riemannian manifold ${\cal N}$ the local  coordinate representation
of which
\begin{eqnarray}
{\cal X}: (\rho,z) \mapsto X^{A}
\end{eqnarray}
satisfies the equations  (\ref{RFE1}) - (\ref{RFE5}) derived from the functional (\ref{FUNCTIONAL}). It is well known that ${\cal X}$ belongs
to the class of the so-called harmonic maps.

The metric $G_{AB}$ can be presented in more explicit form as follows. We consider the matrix ${\mathbb {H}}=h^{-1/2}H \in SL(2,\mr)$. $H$ is positive semi-definite
and therefore ${\mathbb H}$  can be written in the form ${\mathbb H}={\mathbb A}^T{\mathbb A}$ where ${\mathbb A}\in SL(2,\mr)/SO(2)$. Parameterizing
${\mathbb A}$ as

\begin{eqnarray}
{\mathbb A}= \left(
               \begin{array}{cc}
                 e^{X/2} & Ye^{X/2} \\
                 0 & e^{-X/2} \\
               \end{array}
             \right)
\end{eqnarray}
we find

\begin{eqnarray}
&&dL^2= \frac{1}{2}\left(dX^2 + e^{2X}dY^2 \right) + \frac{3}{8}h^{-2}dh^2  + \frac{1}{2}h^{-3/2} \left[e^{-X}\omega_1^2 +
+ e^{X}(\omega_2 - Y\omega_1)^2 \right]  \nonumber \\
&& + 2e^{-2\alpha\varphi} h^{-1/2} \left[e^{-X}d\Phi_1^2 + e^{X}(d\Phi_2 - Y d\Phi_1)^2\right]
+ 2e^{2\alpha\varphi} h^{-1}d\Psi^2 + 2d\varphi^2
\end{eqnarray}
where $\omega_I$ are given by (\ref{omega}).

The important point here, as in the four dimensional case,  is the fact that  the Riemannian  manifold $({\cal N},G_{AB})$ is not a symmetric space in the general case of
arbitrary $\alpha$. This can be seen from the fact that

\begin{eqnarray}
\nabla_{E} R_{ABCD}\ne 0
\end{eqnarray}
as one can check. Only in the case $\alpha^2=8/3$,   $({\cal N},G_{AB})$ is  a symmetric space and this corresponds to  5D Einstein-Maxwell-gravity
obtained as a dimensional reduction of the 6D vacuum Einstein gravity.

\section{Uniqueness theorem for the sectors with single rotation}

From the general system of stationary and axisymmetric  Einstein-Maxwell-dilaton equations we can derive self-consistent  subsystems which describe physically interesting sectors in
Einstein-Maxwell-dilaton  gravity.  One such sector was studied in \cite{Y1},\cite{Y2} and the uniqueness theorem, as we already mentioned,  was proven in \cite{HY2}.  In this section   we will consider another
sector of stationary and axisymmetric 5D Einstein-Maxwell-gravity corresponding to black hole spacetimes with single rotation. To be specific we will consider the case $\omega_2=0$.  The consistency of the field equations    (\ref{RFE1}) -- (\ref{RFE5}) requires  $\Phi_2=0$ and the matrix
$H_{IJ}$ to be diagonal in the canonical basis $\eta_I$.  So the sector is defined by

\begin{eqnarray}
\chi_2=0, \,\, \Phi_2=0, \,\, H=diagonal .
\end{eqnarray}

In the sector under consideration the field equations can be obtained from the functional

\begin{eqnarray}
I^{sec}[X^a]= \int_{{\hat M}}d^2x\sqrt{-{\hat g}} {\hat g}^{ab}G^{sec}_{AB}(X^C){\hat D}_aX^{A} {\hat D}_bX^{B}
\end{eqnarray}
where
\begin{eqnarray}
&&d{L_{sec}^2}=  G^{sec}_{AB} \,dX^AdX^B= \frac{1}{4}Tr\left(H^{-1} dHH^{-1}dH\right) + \frac{1}{4}h^{-2}dh^2  \nonumber \\ && +
\frac{1}{2}h^{-1}H^{11}\left(d\chi_1 + 2\Phi_1 d\Psi - 2 \Psi d\Phi_1\right)^2   + 2e^{-2\alpha\varphi} H^{11}d\Phi_{1}^2
 + 2h^{-1}e^{2\alpha\varphi} d\Psi^2 +2 d\varphi^2  .\nonumber
\end{eqnarray}
is a strictly positive definite metric on the 6-dimensional manifold ${\cal N}^{sec}=\{(H_{11}, H_{22}, \chi_1, \Phi_1, \Psi, \varphi)\in \mr^6;  \,h=H_{11}H_{22}>0\}$.
In the second parameterization we have

\begin{eqnarray}
&&d{L_{sec}^2}= \frac{1}{2}dX^2 + \frac{3}{8}h^{-2}dh^2  + \frac{1}{2}h^{-3/2} e^{-X} (d\chi_1 + 2\Phi_1d\Psi - 2\Psi d\Phi_1)^2  \nonumber \\
&& + 2e^{-2\alpha\varphi} h^{-1/2} e^{-X}d\Phi_1^2
+ 2e^{2\alpha\varphi} h^{-1}d\Psi^2 + 2d\varphi^2
\end{eqnarray}
and ${\cal N}^{sec}=\{(X, h, \chi_1, \Phi_1, \Psi, \varphi)\in \mr^6; h>0\}$.

Now we can formulate the classification theorem. Let us note that the formulation of the classification theorem differs in some points from the formulation of the similar
theorem in \cite{HY2}. More precisely we formulate the classification theorem in terms of the magnetic fluxes. This step is also necessary for the generalization in the next section.

\medskip
\noindent
{\bf Classification Theorem\footnote{Without loss of generality we put $\lim_{\infty}\varphi=0$}:} {\it In the considered sector of 5D Einstein-Maxwell-dilaton gravity there can be at most only one stationary, asymptotically flat
black hole spacetime  satisfying  the technical assumptions stated  in Sec. 2 for a given interval structure $\{l(I_{j}), {\bf a}(I_j)\}$, given angular momentum $J_1$,
given electric charge $Q_{E}$, given magnetic charges $Q_{M}[\mathcal{C}_k]$  for the 2-cycles $\mathcal{C}_k$, given left magnetic flux ${\mathfrak{F}_L}$
and for dilaton coupling parameter satisfying $0\le \alpha^2\le 8/3$.}

\medskip
\noindent

{\bf Remark1:} The definition assumptions for the sector under consideration imply that ${\bf a}(I_j) = (1,0)$ or $(0,1)$, and according to the topology
theorem of Sec. II the horizon topology is $ S^3$ or $S^1 \times S^2$.

\medskip
\noindent

{\bf Remark2:} Without loss of generality we orient the interval structure so that the semi-infinite interval corresponding to the Killing vector $\eta_1$ to be on the right
side of the horizon interval.  The 2-cycles $\mathcal{C}_k$ and the associated magnetic charges $Q_{M}[\mathcal{C}_k]$ are defined as follows.
We consider  a simple curve ${\hat \gamma}_k$ that starts on a $\eta_1$-axis interval and ends on the next  $\eta_1$-axis interval with neither   $\eta_1$-axis interval nor the horizon
interval in between. The closed 2-surface $\mathcal{C}_k$ with  topology of a 2-sphere is  generated first by lifting ${\hat \gamma}_k$ to a
curve $\gamma_k$ in $\langle\langle M\rangle\rangle$ and then acting with the isometries generated by $\eta_2$. Then the magnetic charges are given by

\begin{eqnarray}
Q_{M}[\mathcal{C}_k]=\int_{\mathcal{C}_k}F .
\end{eqnarray}

The left magnetic flux ${\mathfrak{F}_L}$ is defined as follows. We consider an arbitrary simple curve ${\hat \gamma}_{-}$ in the interior of ${\hat M}$
starting from the intersection point, $z_1$, of the semi-infinite interval of $\eta_2$ with the first finite interval (from left to right) and going to infinity with $z\to -\infty$.
Then the left magnetic flux is given by

\begin{eqnarray}
{\mathfrak{F}_L}=\int_{D_{L}}F
\end{eqnarray}
where $D_{L}$ is a 2-surface with disk topology generated first by  lifting ${\hat \gamma}_{-}$ to a curve $\gamma_{-}$ in $\langle\langle M\rangle\rangle$
and then acting with the isometries generated by $\eta_1$.

\medskip
\noindent

Before presenting the very proof of the theorem we need two intermediate results related to important   properties of the Riemannian manifold  $({\cal N}^{sec}, G^{sec}_{AB})$.

\medskip
\noindent
{\bf Lemma:} {\it The Riemannian manifold $({\cal N}^{sec}, G^{sec}_{AB})$ is geodesically complete for any $\alpha$.}

\medskip
\noindent

{Proof:}  Let $s\mapsto \gamma(s)$ be an affinely parameterized geodesic,

\begin{eqnarray}
\gamma(s)=\left(X(s), h(s), \chi_1(s),\Phi_1(s),\Psi(s),\varphi(s)\right).
\end{eqnarray}

Then $G^{sec}(\dot \gamma,\dot \gamma)=C>0$ is a constant of motion, i.e.

\begin{eqnarray}
&&\frac{1}{2} {\dot X}^2 + \frac{3}{8} \left(\frac{\dot h}{h}\right)^2 +
\frac{1}{2}h^{-3/2} e^{-X} (\dot\chi_1 + 2\Phi_1{\dot \Psi} - 2\Psi {\dot\Phi_1})^2  \nonumber \\
&& + 2e^{-2\alpha\varphi} h^{-1/2} e^{-X}{\dot\Phi_1}^2
+ 2e^{2\alpha\varphi} h^{-1}{\dot \Psi}^2 + 2{\dot\varphi}^2=C.
\end{eqnarray}
Therefore  we have

\begin{eqnarray}
{\dot X}^2\le 2C, \;\;\;  \left(\frac{\dot h}{h}\right)^2\le \frac{8}{3} C, \;\;\; {\dot\varphi}^2\le \frac{1}{2}C
\end{eqnarray}
and
\begin{eqnarray}
{\dot\Phi_1}^2 \le \frac{1}{2}Ce^{2\alpha\varphi}h^{1/2} e^{X}, \;\;\; {\dot \Psi}^2\le \frac{1}{2}Ce^{-2\alpha\varphi}h, \;\;\;
(\dot\chi_1 + 2\Phi_1{\dot \Psi} - 2\Psi {\dot\Phi_1})^2\le 2C h^{3/2} e^{X}.
\end{eqnarray}
These inequalities show that any geodesic can  be extended to arbitrary value of the affine parameter.

\medskip
\noindent
{\bf Lemma:} {\it The Riemannian manifold $({\cal N}^{sec}, G^{sec}_{AB})$   is manifold with nonpositive Riemann curvature operator
for a dilaton coupling parameter  $\alpha$ satisfying  $0\le \alpha^2\le 8/3$.}

\medskip
\noindent

{Proof:} We regard the Riemann curvature tensor as an operator $\hat {\cal R}: \Lambda^2T({\cal N}^{sec})\rightarrow \Lambda^2T({\cal N}^{sec})$ (
or as an operator $\hat {\cal R}: \Lambda^2T^{*}({\cal N}^{sec})\rightarrow \Lambda^2T^{*}({\cal N}^{sec}) $) which is self-adjoint (symmetric)
with respect to the naturally induced metric on $\Lambda^2T({\cal N}^{sec})$ (or on $\Lambda^2T^{*}({\cal N}^{sec})$). By direct computation of the eigenvalues $\lambda_i$ of the matrix of the operator $\hat {\cal R}$  we find that

\begin{eqnarray}
 \lambda_1=\lambda_2=...=\lambda_{11}=0,\; \;\lambda_{12}=\lambda_{13}=-\frac{7}{3} -  \alpha^2, \\
 \lambda_{14, 15}= - \frac{23}{6} + \frac{\alpha^2}{2}  \pm  \frac{\sqrt{145  + 6\alpha^2 + 9\alpha^4}}{6}. \nonumber
\end{eqnarray}

For  $0\le \alpha^2\le 8/3$ all eigenvalues are nonpositive and therefore the Riemann operator is nonpositive. As an immediate consequence we have that
$({\cal N}^{sec}, G^{sec}_{AB})$  is a manifold with nonpositive sectional curvature.

\medskip
\noindent

{\bf Proof of the Theorem:}  Consider two solutions $(M, g, F, \varphi)$ and $({\tilde M}, {\tilde g}, {\tilde F}, {\tilde \varphi})$.
We use the same "tilde" notation to distinguish any quantities associated with the two solutions.
Since both solutions have the same interval structure,  we can identify the orbit spaces ${\hat M}$ and ${\hat {\tilde M}}$. Moreover we can identify $\langle\langle M\rangle\rangle$ and $\langle\langle{\tilde M}\rangle\rangle$ as manifolds with $\mr \times U(1)^2$-action since they can be uniquely reconstructed from the orbit space. We may therefore
assume that $\langle\langle M\rangle\rangle=\langle\langle{\tilde M}\rangle\rangle$ and that $\xi={\tilde \xi}$, $\eta_1={\tilde \eta}_1$, $\eta_2={\tilde \eta}_2$. We may also assume that $\rho={\tilde \rho}$  and $z={\tilde z}$. As a consequence of these identifications, $(g, F, \varphi)$  and $({\tilde g}, {\tilde F}, {\tilde \varphi})$  may be considered as being defined on the same manifold.

Further we consider two harmonic maps  (\ref{harmonicmapp}) ${\cal X}: {\hat M} \mapsto {\cal N}^{sec}$ and ${\tilde {\cal X}}: {\hat M} \mapsto {\cal N}^{sec}$, and
a smooth  homotopy

\begin{eqnarray}
{\cal T}: {\hat M}\times [0,1] \mapsto {\cal N}^{sec}
\end{eqnarray}
 so that
\begin{eqnarray}
{\cal T}(\tau=0)={\cal X}, \;\;\;\; \;{\cal T}(\tau=1)={\tilde {\cal X}}
\end{eqnarray}
where  $0\le \tau\le 1$ is the homotopy parameter. Rephrasing in local terms we consider  two solutions $X^{A}(\rho,z)$ and ${\tilde X}^A(\rho,z)$  and
a smooth homotopy ${\cal T}: {\hat M}\times [0,1] \mapsto {\cal N}^{sec}$ such that
\begin{eqnarray}
{\cal T}^{A}(\rho,z;\tau=0)=X^A(\rho,z) ,\;\;\;\;\;\;   {\cal T}^{A}(\rho,z;\tau=1)={\tilde X}^{A}(\rho,z)
\end{eqnarray}
for each point $(\rho,z)\in {\hat M}$.    As a further requirement we impose that the
curves $[0,1]\mapsto {\cal N}^{sec}$  be geodesic which in local coordinates means that

\begin{eqnarray}
\frac{dS^A}{d\tau} + \Gamma^{A}_{B\,C}S^{B}S^{C}=0
\end{eqnarray}
where
\begin{eqnarray}
S^A=\frac{d{\cal T}^A}{d\tau}
\end{eqnarray}
is the tangent vector along the curves and $\Gamma^{A}_{B\,C}$ are components of the Levi-Civita connection on ${\cal N}^{sec}$.

The existence and uniqueness of the geodesic homotopy follow from well known results in Riemannian geometry \cite{Postnikov} since
the Riemannian  manifold $({\cal N}^{sec}, G^{sec}_{AB})$ is geodesically complete, simply connected and with nonpositive sectional curvature.

The length of the geodesics will be denoted by $S$, i.e.

\begin{eqnarray}
S= \int^{1}_{0} d\tau\, G_{AB}S^A S^B
\end{eqnarray}
with the $\tau$-independent normalization condition $S^A S_A=S^2$.

Now we write the  Bunting identity \cite{Bunting},\cite{Carter1}

\begin{eqnarray}\label{Bidentity}
{\hat D}_a\left(\rho S{\hat D}^a S \right)= \rho \int^{1}_{0} d\tau \left({\hat \nabla}^a S_A {\hat \nabla }_a S^A
 - R_{ABCD} S^A {\hat \nabla}_a{\cal T}^B  S^C {\hat \nabla}^a {\cal T}^D  \right).
\end{eqnarray}
Here ${\hat \nabla}_a$ is  the induced connection along the harmonic map, i.e.

\begin{eqnarray}
{\hat \nabla}_a V^A  = \partial_a{\cal T}^C \nabla_{C} V^A= \partial_a V^A +    \partial_a {\cal T}^C \Gamma^{A}_{CB} V^{B}.
\end{eqnarray}

Since $\rho\ge 0$, ${\cal N}^{sec}$ has positive definite metric and nonpositive sectional curvature for $0\le \alpha^2\le 8/3$ we conclude that

\begin{eqnarray}\label{SINEQ}
{\hat D}_a\left(\rho S{\hat D}^a S \right)\ge 0 .
\end{eqnarray}

At this stage it is convenient to view $\rho$ and $z$ as cylindrical coordinates in an auxiliary space $\mr^3$ consisting of the points $X=(\rho\cos\phi,\rho\sin\phi,z)$.
It is also convenient to  view the geodesic distance $S$ as an axially symmetric function on $\mr^3\backslash\{z-axis\}$.  Eq.(\ref{SINEQ}) can then be written as

\begin{eqnarray}
\Delta S^2\ge 0,
\end{eqnarray}
where $\Delta$ is the ordinary Laplacian on $\mr^3$.

The requirement  $\Delta S^2\ge 0$ imposes very strong constraints on $S^2$ (respectively on $S$). If one can show that $S$ is globally bounded on $\mr^3$ including the
$z$-axis and vanishes at infinity then $S$ must vanish everywhere \cite{W1,W2}. So our next step is to show that $S$ is indeed globally bounded. More precisely it is sufficient
to prove that $S$ is bounded in the following cases: (i) in an open neighborhood of the open interval $I_{H}$ representing the horizon,
(ii) at infinity, (iii) in an open
neighborhood of each open interval $I_j$ corresponding to a rotational axis,  (iv) in open neighborhoods of the corners.

\medskip
\noindent

(i) On the open interval of the horizon the matrix $H$ is invertible so $S$ is bounded there.

\medskip
\noindent

In order to show that $S$ is bounded on the intervals $I_j$ corresponding to a rotational axis  and at infinity we consider a piecewise differentiable curve $\gamma_{*}:[t_0, t_*]\to {\cal N}^{sec}$ which joins the points $\{X^{A}\}$ and $\{{\tilde X}^{A}\}$ on ${\cal N}^{sec}$ and which is  defined as follows. Take a partition $t_0<t_1<t_2<t_3<t_5<t_5<t_6=t_*$ such that the restrictions ${\gamma_*}_{[t_{A-1},t_{A}]}$ are differentiable curves explicitly given by

\begin{eqnarray}
{\gamma_*}_{[t_{A-1},t_{A}]}=\{\gamma^{\,A}_*(t)=X^A + ({\tilde X}^A - X^A)\frac{t_A - t}{t_{A}-t_{A-1}}, \;\; \gamma^{\,B\ne A}_*(t) = const= \gamma^{\,B\ne A}_*(t_{A-1})\}.
\end{eqnarray}

The length $S_*$ of $\gamma_*$ can be  easily found and the result is\footnote{Here we use the potential $\sigma_1$ instead of $\chi_1$.}

\begin{eqnarray}\label{SSTAR}
&&S_*= \frac{1}{\sqrt{2}} | \ln\left(\frac{{\tilde H}_{11}}{H_{11}}\right)| +  \frac{1}{\sqrt{2}} | \ln\left(\frac{{\tilde H}_{22}}{H_{22}}\right)| +
\frac{1}{\sqrt{2}} \frac{|{\tilde \sigma_1} - \sigma_1 |}{{\tilde H}_{11}\sqrt{{\tilde H}_{22}}} \\
&&+  \sqrt{2}e^{-\alpha\varphi } \frac{|{\tilde \Phi}_1 - \Phi_1|}{\sqrt{{\tilde H}_{11}}} +
 \sqrt{8{\tilde \Phi_1}^2 + 2e^{2\alpha\varphi}{\tilde H}_{11}} \frac{|{\tilde \Psi} - \Psi|}{{\tilde H}_{11}\sqrt{{\tilde H}_{22}}} + \sqrt{2}|{\tilde \varphi} - \varphi| .
\nonumber
\end{eqnarray}

Due to the global properties of ${\cal N}^{sec}$ and the fact the $S$ is a geodesic distance we must have $S\le S_*$. Therefore it is sufficient to show that $S_*$ is globally bounded.

\medskip
\noindent

(ii) Near infinity the asymptotic form of the metric is given by

\begin{eqnarray}\label{METRICASYMPT}
&&ds^2\approx  -\left(1 - \frac{8M}{3\pi r^2} + {{\cal O}(r^{-4})}\right)dt^2 - \left(\frac{8J_1\sin^2\theta}{\pi r^2} + {{\cal O}(r^{-4})} \right)dt d\phi_1
 - \left(\frac{8J_2\cos^2\theta}{\pi r^2} + {{\cal O}(r^{-4})} \right)dt d\phi_2
 \nonumber\\ &&
+ \left(1 + {\cal O}(r^{-2})\right)\left(dr^2  + r^2d\theta^2\right) +
r^2 \sin^2\theta \left(1+ {\cal O}(r^{-2})\right)d\phi_1^2+ r^2\cos^2\theta \left(1 + {\cal O}(r^{-2})\right) d\phi_2^2  \\
&&+ {\cal O}(\frac{\sin^2\theta\cos^2\theta}{r^2})d\phi_1 d\phi_2 \nonumber
\end{eqnarray}
where $M$ and $J_I$ are the mass and angular momenta defined by

\begin{eqnarray}
M= \frac{3}{32\pi}\int_{S_{\infty}^3}\star\, d\xi, \\
J_{I}= \frac{1}{16\pi}\int_{S_{\infty}^3} \star\, d\eta_I.
\end{eqnarray}
In writing the asymptotic metric we have used the asymptotic coordinates $r$ and  $\theta$ given by

\begin{eqnarray}
\rho= \frac{r^2}{2}\sin(2\theta), \;\;\;\; z=\frac{r^2}{2}\cos(2\theta).
\end{eqnarray}
In the case under consideration, by definition  we have $J_2=0$

The asymptotic behavior of the other potentials is as follows:

\begin{eqnarray}
&&\Psi= - \frac{Q_E}{2}\left(\cos^2\theta -\frac{1}{2}\right)  + {\cal O}\left(\frac{\sin^2\theta\cos^2\theta}{r^2} \right) ,\\
&&\chi_{1}=\frac{4J_1}{\pi} \left(\sin^4\theta - \frac{1}{2}\right) + {\cal O}\left(\frac{\sin^2\theta\cos^2\theta}{r^2} \right), \\
&&\Phi_1= {\cal O}\left(\frac{\sin^2\theta}{r^2}\right),\\
&&\varphi = {\cal O}\left(\frac{1}{r^2}\right),
\end{eqnarray}
where $Q_E$ is the electric charge defined by

\begin{eqnarray}
Q_E= \frac{1}{2\pi^2}\int_{S_{\infty}^3} e^{-2\alpha\varphi}\star F .
\end{eqnarray}

Taking into account the above asymptotics we find that $S_*={\cal O}(r^{-2})$ near infinity. This shows that  $S_*$ and therefore also $S$ tends to zero near infinity uniformly in
$\theta$ including the axis.

\medskip
\noindent

(iii) Let us first consider the behavior of the potentials $\Psi$ and $\sigma_1$ on the $z$-axis outside the horizon. By the definition of $\Psi$ we have that $\Psi$ is a constant on the $z$-axis outside the horizon. The difference between the constant value of $\Psi$ on the $z$-axis  right and left to the horizon interval is given by

\begin{eqnarray}
\Psi(z_{h+1},\rho=0) - \Psi(z_{h},\rho=0)= \int_{I_H}d\Psi= -\frac{1}{4\pi^2} \int_{H}e^{-2\alpha\varphi}\star F= -\frac{Q_E}{2}
\end{eqnarray}
and similar expression for the tilde solutions, namely

\begin{eqnarray}
{\tilde \Psi}(z_{h+1},\rho=0) - {\tilde \Psi}(z_{h},\rho=0)=  -\frac{{\tilde Q}_E}{2}=  -\frac{Q_E}{2}.
\end{eqnarray}
Therefore we obtain that ${\tilde \Psi}-\Psi= const$ on the  $z$-axis outside the horizon. Since ${\tilde \Psi}=\Psi$ at infinity we find that ${\tilde \Psi}=\Psi$
everywhere on  the $z$-axis outside the horizon. Using again the fact that $d\Psi$ vanishes on   the $z$-axis outside the horizon we conclude that
${\tilde \Psi}- \Psi= {\cal O}(\rho^2)$ near  the $z$-axis outside the horizon.

Consider now the behavior of $\sigma_1$. By definition $\sigma_1$ is constant on  the $z$-axis outside the horizon. The difference between the constant value of $\sigma_1$ on the $z$-axis  right and left to the horizon interval is

\begin{eqnarray}
{\sigma_1}(z_{h+1},\rho=0) - {\sigma_1}(z_{h},\rho=0)=  \int_{I_H}d\sigma_1=\int_{\partial {\hat M}\bigcup \infty} d\sigma_1 - \int_{\infty}d\sigma_1= \nonumber\\
- \int_{\infty}d\sigma_1=
 -\int_{\infty}\omega_1 + 4\int_{\infty}\Phi_1d\Psi =  -\int_{\infty}\omega_1 = - \frac{1}{4\pi^2} \int_{S_{\infty}^3}\star d\eta_1= -\frac{4J_1}{\pi}
\end{eqnarray}
where we have taken into account that $\lim_{\infty}\Phi_{1}=0$. We also have analogous expression for the tilde solution
\begin{eqnarray}
{\tilde \sigma_1}(z_{h+1},\rho=0) - {\tilde \sigma_1}(z_{h},\rho=0)=  -\frac{4{\tilde J}_1}{\pi}= -\frac{4J_1}{\pi}.
\end{eqnarray}
Therefore we obtain that ${\tilde \sigma_1}- \sigma_{1}=const$ on the $z$-axis outside the horizon. Since ${\tilde \sigma_1}= \sigma_{1}$ at infinity we conclude that
${\tilde \sigma_1}= \sigma_{1}$  everywhere on the $z$-axis outside the horizon which together with the fact that $d\sigma_1=0$ there shows that
${\tilde \sigma_1} - \sigma_{1}={\cal O}(\rho^2)$ near the $z$-axis outside the horizon.

It remains to consider the behavior of the potential $\Phi_1$ on the axes of $\eta_1$. From the definition $d\Phi_1=i_{\eta_1}F$ it follows that $\Phi_1$
is constant on these axes. The difference of the constant value on two neighbor $\eta_1$-axes is

\begin{eqnarray}\label{MAGCHARGE}
\Phi^{i+1}_1- \Phi^{i-1}_{1}= \int_{{\hat \gamma}_i}d\Phi_1= \frac{1}{2\pi}\int_{\mathcal{C}_i}F=\frac{1}{2\pi}Q_M[\mathcal{C}_i]
\end{eqnarray}
and a similar expression for the tilde solution

\begin{eqnarray}
{\tilde \Phi}^{i+1}_1- {\tilde \Phi}^{i-1}_{1}=\frac{1}{2\pi}{\tilde Q}_M[\mathcal{C}_i]
\end{eqnarray}

By direct computation we can also find the left magnetic flux

\begin{eqnarray}\label{MAGFLUX}
{\mathfrak{F}_L}= 2\pi \Phi_{1}(z_1), \;\;\; \; {\tilde {\mathfrak{F}}}_L= 2\pi {\tilde \Phi}_{1}(z_1).
\end{eqnarray}

Let us enumerate the intervals to the left of the horizon interval by $I_a, a=0,1,2,....$ as $a=0$ corresponds to the semi-infinite interval of $\eta_2$.
In this case the odd $a$ describe the $\eta_1$-axis intervals while the even $a$ describe the $\eta_2$-axis intervals. Using   (\ref{MAGCHARGE}) and
(\ref{MAGFLUX}) it is not difficult one to show that

\begin{eqnarray}
\Phi^{2k+1}_{1}= \frac{1}{2\pi}\sum^{k}_{l=1} Q_{M}[\mathcal{C}_{2l}] + \frac{1}{2\pi} {\mathfrak{F}_L},\;\;\;
\end{eqnarray}
where $\Phi^{2k+1}_{1}$ is the value of  the potential $\Phi_1$ on the $2k+1$-th axis of $\eta_1$. In the same way we can enumerate the intervals right to the horizon interval
by $I_b$ where $b=1$ corresponds to the semi-infinite interval of $\eta_1$.  Then,  for the value  $\Phi^{2n+1}_1$ of $\Phi_1$ on $2n+1$-th interval of the $\eta_1$-axis  we find

\begin{eqnarray}
\Phi^{2n+1}=  \frac{1}{2\pi}\sum^{n}_{l=1} Q_{M}[\mathcal{C}_{2l}] \;\;\;\;\; (\Phi^{1}_1=0)
\end{eqnarray}

Taking into account that both solutions have the same magnetic charges and the same left magnetic flux we conclude that
${\tilde \Phi}_1 = {\Phi}_1$ on the axes of $\eta_1$. This, together with the fact that $d\Phi_1=0$ on the $\eta_1$-axes show that
${\tilde \Phi}_1 - {\Phi}_1={\cal O}(\rho^2)$ near the $\eta_1$-axes.

Let us first focus on the  open intervals corresponding to $\eta_1=0$ and $\eta_2\ne 0$ i.e. the intervals with  ${\bf a}=(1,0)$. For such intervals  $H_{22}\ne0$,
 ${\tilde H}_{22}\ne0$   and $H_{11}={\cal O}(\rho^2)$, ${\tilde H}_{11}={\cal O}(\rho^2)$. This shows that the first and second terms of (\ref{SSTAR}) are
 bounded. Further using the fact that near the open intervals corresponding to  $\eta_1=0$ and $\eta_2\ne 0$ we have ${\tilde \sigma_1}- \sigma_1= {\cal O}(\rho^2)$,
 ${\tilde \Phi_1}- \Phi_1= {\cal O}(\rho^2)$ and ${\tilde \Psi}- \Psi= {\cal O}(\rho^2)$ we conclude that the third, the forth and the fifth terms  in (\ref{SSTAR}) are
 bounded. Obviously the last term is also bounded.

Now let us consider the open intervals corresponding $\eta_2=0$ and $\eta_1\ne 0$ i.e. open intervals with ${\bf a}=(0,1)$.  For such intervals we have
$H_{22}={\cal O}(\rho^2)$ and $H_{11}\ne 0$. This shows that the first and second terms in (\ref{SSTAR})  are bounded. The forth and the sixth terms are
obviously bounded. Taking into account that near intervals under consideration  ${\tilde \sigma_1}- \sigma_1= {\cal O}(\rho^2)$  and ${\tilde \Psi}- \Psi= {\cal O}(\rho^2)$ we conclude that the third and the fifth terms are bounded, too.

 \medskip
\noindent

(iv) The continuity argument shows that $S_*$ is bounded  in open neighborhoods of the corners.

 \medskip
\noindent

We have shown that $S_*$ and consequently  also $S$ is globally bounded on $\mr^3$ including the $z$-axis and vanishes at infinity.
Therefore $S$ vanishes everywhere and this completes the proof.

\section{Uniqueness theorem in the general case}

In this section  we consider the uniqueness problem in the general case when the electromagnetic field is fully excited and both angular momenta $J_1$ and $J_2$ are
non-zero.  In the general case however we can prove  the uniqueness  theorem only by imposing very tight restriction on the dilaton coupling parameter $\alpha$.
The method of the proof however is logically the same as in the previous section with some technical complications.

\medskip
\noindent

{\bf Lemma:} {\it The Riemannian manifold $({\cal N}, G_{AB})$ is geodesically complete for any $\alpha$.}

 \medskip
\noindent

The proof is the same as in the previous section and we do not present it here.

The next lemma follows from the direct computation of the eigenvalues of the Riemann curvature operator.

 \medskip
\noindent

{\bf Lemma:} {\it The Riemann manifold  $({\cal N}, G_{AB})$ is manifold with nonpositive Riemann curvature operator only for $\alpha^2=8/3$.}

 \medskip
\noindent

Now we are at  the position to formulate the theorem, namely:
 \medskip
\noindent

{\bf Uniqueness Theorem:} {\it There can be only one stationary, asymptotically flat black hole spacetime satisfying the 5D Einstein-Maxwell-dilaton field equations
and the technical assumptions stated in Sec. 2 for a given interval structure $\{l(I_j), {\bf a}(I_j)\}$, given angular momenta $J_1$ and $J_2$, given electric charge
$Q_{E}$, given magnetic charges $Q_{M}[\mathcal{C}_k]$  for the 2-cycles $\mathcal{C}_k$, given left  and right magnetic fluxes ${\mathfrak{F}_L}$ and
${\mathfrak{F}_R}$, and for dilaton coupling parameter $\alpha^2= 8/3$.}

\medskip
\noindent
{\bf Remark:} The 2-cycles $\mathcal{C}_k$ are associated with the finite intervals different from the horizon interval and are defined as follows. Consider
a finite interval $[z_k,z_{k+1}]$ different from the horizon and the linear combination $\eta^{(k)}=a^{k}_1\eta_1 + a^{k}_2\eta_2$ that vanishes on it. Let $\hat \gamma_{k}$
is an arbitrary simple curve
in the interior of $\hat M$ that starts on $z_k$ and ends on $z_{k+1}$.  The closed 2-surface $\mathcal{C}_k$ with  topology of a 2-sphere is  generated first by lifting ${\hat \gamma}_k$ to a
curve $\gamma_k$ in $\langle\langle M\rangle\rangle$ and then acting with the isometries generated
by $\eta^{(k-1)}=a^{k-1}_1\eta_1 + a^{k-1}_2\eta_2$ or $\eta^{k+1}=a^{k+1}_1\eta_1 + a^{k+1}_2\eta_2$ where $\eta^{(k-1)}$ and $\eta^{(k+1)}$ are
linear combinations that vanish on the interval (not necessary finite but different from the horizon)  left to or  right to  $[z_k,z_{k+1}]$ . The magnetic charge is then given by the same formula
as in the previous section.

The left (right)  magnetic flux ${\mathfrak{F}_L}$ (${\mathfrak{F}_R}$)  is defined as follows. We consider an arbitrary curve ${\hat \gamma}_{-}$ (${\hat \gamma}_{+}$)
in the interior of ${\hat M}$ starting from the intersection point, $z_1$ ($z_{N+1}$) (see (\ref{zintervals})) and  going to infinity with $z\to -\infty$ ($z\to +\infty$).
Then the left (right) magnetic  flux is given by

\begin{eqnarray}
{\mathfrak{F}_{L,R}}=\int_{D_{L,R}}F
\end{eqnarray}
where $D_{L}$ ($D_R$) is a 2-surface with disk topology generated first by  lifting ${\hat \gamma}_{-}$  (${\hat \gamma}_{+}$) to a curve
$\gamma_{-}$ ($\gamma_{+}$ ) in $\langle\langle M\rangle\rangle$ and then acting with the isometries generated by $\eta_1$ ($\eta_2$).

 \medskip
\noindent

{\bf Proof:}  The idea of the proof is the same as in the previous section and that is why we shall give only the basic steps. Since the Riemann curvature operator is
nonpositive for $\alpha^2= 8/3$ the same is true for the sectional curvature and therefore the geodesic distance satisfies

\begin{eqnarray}
\Delta S^2 \ge 0
\end{eqnarray}
where $\Delta$, as in the previous section, is the ordinary Laplacian on $\mr^3$. We have to show that $S$ in globally bounded on  $\mr^3$ including the $z$-axis.

 \medskip
\noindent

(i) On the open interval of the horizon the matrix $H$ is invertible so $S$ is bounded there.

 \medskip
\noindent

As in the previous section in order to show that $S$ is bounded at infinity and on the intervals corresponding to a rotational axis we consider an auxiliary curve
joining the points ${\tilde X}^A$ and $X^A$ on ${\cal N}$ with a length $S_*$ given by
\begin{eqnarray}\label{SSTARGEN}
S_{*} &=&\frac{1}{\sqrt{2}} \left|\ln\left(\frac{{\tilde h}}{h}\right)\right|
+ \frac{1}{\sqrt{2}}\sqrt{{\tilde h}^{-1}{\tilde H}^{IJ}\left({\tilde \sigma}_{I}- \sigma_I\right)\left({\tilde \sigma}_{J}- \sigma_J\right)}
+ \sqrt{2}e^{-\alpha\varphi} \sqrt{{\tilde H}^{IJ}\left({\tilde \Phi}_{I}- \Phi_I\right)\left({\tilde \Phi}_{J}- \Phi_J\right)} \nonumber \\
 &&+ \sqrt{2} \sqrt{{\tilde h}^{-1}\left[4{\tilde H}^{IJ}{\tilde \Phi}_{I}{\tilde \Phi}_{J} + e^{-2\alpha\varphi}\right]}\left|{\tilde \Psi} - \Psi\right| +
\sqrt{2}\left|{\tilde \varphi} -\varphi\right| .
\end{eqnarray}
Since $S\le S_*$ it is sufficient to show that $S_*$ is globally bounded.

 \medskip
\noindent

(ii) Near infinity we have the following asymptotic behavior

\begin{eqnarray}
&&\Psi=  - \frac{Q_E}{2}\left(\cos^2\theta -\frac{1}{2}\right)  + {\cal O}\left(\frac{\sin^2\theta\cos^2\theta}{r^2} \right) ,\\
&&\chi_{1}=\frac{4J_{1}}{\pi} \left(\sin^4\theta - \frac{1}{2}\right) + {\cal O}\left(\frac{\sin^2\theta\cos^2\theta}{r^2} \right), \\
&&\chi_{2}=\frac{4J_{2}}{\pi} \left(\cos^4\theta - \frac{1}{2}\right) + {\cal O}\left(\frac{\sin^2\theta\cos^2\theta}{r^2} \right), \\
&&\Phi_1= {\cal O}\left(\frac{\sin^2\theta}{r^2}\right),\\
&&\Phi_2= {\cal O}\left(\frac{\cos^2\theta}{r^2}\right),\\
&&\varphi = {\cal O}\left(\frac{1}{r^2}\right),
\end{eqnarray}

Taking into account these asymptotics and the asymptotic of the spacetime metric (\ref{METRICASYMPT}) we find that $S_*={\cal O}(r^{-2})$.
This shows that  $S_*$ and therefore also $S$ tends to zero near infinity uniformly in $\theta$ including the $z$-axis.

 \medskip
\noindent

(iii) In the same way as in the previous section one can show that ${\tilde \Psi} - \Psi={\cal O}(\rho^2)$ and ${\tilde \sigma}_{I}-\sigma_{I}={\cal O}(\rho^2)$
near the $z$-axis outside the horizon. Let us now consider the behavior of the potentials $\Phi_I$ on the $z$-axis outside the horizon.  Fix an interval (finite or semi-infinite)
$I_{k}$ with an associated vector ${\bf a}^{k}=(a^{k}_1,a^{k}_2)\ne (0,0)$ and let us denote by $Y^{k}$ the linear combination

\begin{eqnarray}
Y^{k}= a^{k}_1 \Phi_1 + a^{k}_2\Phi_2.
\end{eqnarray}
Since the linear combination $a^{k}_1 \eta_1 + a^{k}_2\eta_2$ vanishes on $I_{k}$ it follows that $Y^{k}$ is constant on $I_{k}$. When $I_k$ is a finite interval we can compute the
magnetic charge associated with it, namely

\begin{eqnarray}\label{MAGCHARGE1}
 \frac{1}{2\pi} Q_{M}[\mathcal{C}_k] = Y^{k+1} - Y^{k-1} - \det({\bf a}^{k-1},{\bf a}^{k+1}) Y^{k}.
\end{eqnarray}

Consider now the intervals left to the horizon interval. The expression (\ref{MAGCHARGE1}) can then be  rewritten as a recurrent dependence

\begin{eqnarray}\label{RECURRENT}
Y^{k+1} = Y^{k-1}  + \det({\bf a}^{k-1},{\bf a}^{k+1}) Y^{k} + \frac{1}{2\pi}  Q_{M}[\mathcal{C}_k]
\end{eqnarray}
where $k=0$ corresponds to the leftmost semi-infinite interval. Further taking into account that $Y^{0}=\Phi_{2}(-\infty)=0$ and that

\begin{eqnarray}
{\mathfrak{F}_{L}}= 2\pi \Phi_{1}(z_1)= 2\pi Y^{1}(z_1)
\end{eqnarray}
we can solve the recurrent dependence (\ref{RECURRENT})

\begin{eqnarray}
&&Y^{2}= \frac{1}{2\pi}  \det({\bf a}^{0},{\bf a}^{2}){\mathfrak{F}_{L}} +  \frac{1}{2\pi}Q_{M}[\mathcal{C}_1] , \nonumber \\
&&Y^{3}= \frac{1}{2\pi} {\mathfrak{F}_{L}} + \det({\bf a}^{1},{\bf a}^{3})\left(\frac{1}{2\pi}  \det({\bf a}^{0},{\bf a}^{2}){\mathfrak{F}_{L}} + \frac{1}{2\pi} Q_{M}[\mathcal{C}_1]\right)
+   \frac{1}{2\pi}Q_{M}[\mathcal{C}_2] \\
&&Y^4 =... \nonumber\\
&&Y^{5}=... \nonumber\\
&& ...  \nonumber
\end{eqnarray}
Therefore all $Y^{k}$ associated with the intervals left to the horizon are fully determined by the magnetic charges and the left magnetic flux.
The same can be repeated for the intervals right to the horizon. So we conclude that for intervals (different from the horizon) $Y^{k}$  are fully determined by
the magnetic charges and the left and right magnetic flux.

The whole scheme can be repeated  for the tilde solution. Making use of the fact that both solutions have the same interval structure, the same magnetic charges and the same
magnetic fluxes we conclude that ${\tilde Y}^{k}-Y^{k}=0$ on $I_{k}$. Since $dY^{k}=0$ we find that   ${\tilde Y}^{k}-Y^{k}=O(\rho^2)$ near $I_{k}$.

Now we can show that each term of $S_*$ is uniformly bounded near the axis represented by $I_k$. Let ${\bf a}^{k}$  be the vector associated with the $I_{k}$, i.e.
the vector generating the kernel of the matrix $H_{IJ}$. One can find a matrix $B\in SL(2,\mz)$ such that $B\,{\bf a}^{k}=(1,0)$ (see \cite{HY3}). Thus redefining the Killing fields
as $\eta_I\mapsto B^{-1}_{IJ}\eta_J$   we can assume that ${\bf a}^{k}=(1,0)$. In this basis  $H$  takes the  following form near $I_{k}$

\begin{eqnarray}
H=\left(
    \begin{array}{cc}
      \rho^2 e^{2\nu} + {\cal O}(\rho^2) & {\cal O}(\rho^4) \\
       {\cal O}(\rho^2)  & e^{2\mu} + {\cal O}(\rho^2) \\
    \end{array}
  \right)
\end{eqnarray}
where $\nu$ and $\mu$ are regular functions. The same form has also the matrix ${\tilde H}$ associated with  the tilde solution.

From the above expression it follows that the first term in  (\ref{SSTARGEN}) is bounded. The second term behaves as

\begin{eqnarray}
{\tilde h}^{-1}{\tilde H}^{IJ}\left({\tilde \sigma}_{I}- \sigma_I\right)\left({\tilde \sigma}_{J}- \sigma_J\right)=
e^{-4{\tilde \nu}-2{\tilde \mu}} \frac{\left({\tilde \Lambda}^{k} -\Lambda^{k}\right)^2}{\rho^4} + {\cal O}(\rho^2)
\end{eqnarray}
 where $\Lambda^{k}= a^{k}_1\sigma_1 + a^{k}_{2}\sigma_2$. Taking into account that ${\tilde \Lambda}^{k} -\Lambda^{k}={\cal O}(\rho^2)$ we conclude that the
 second term in (\ref{SSTARGEN}) is bounded.

For the third term we have

\begin{eqnarray}
{\tilde H}^{IJ}\left({\tilde \Phi}_{I}- \Phi_I\right)\left({\tilde \Phi}_{J}- \Phi_J\right)=e^{-2{\tilde \nu}} \frac{\left({\tilde Y}^k - Y^k\right)^2}{\rho^2} + {\cal O}(\rho^2)
\end{eqnarray}
which shows that the third term is bounded since ${\tilde Y}^k - Y^k={\cal O}(\rho^2)$.

The behavior of the forth term is
\begin{eqnarray}
 {\tilde h}^{-1}\left[4{\tilde H}^{IJ}{\tilde \Phi}_{I}{\tilde \Phi}_{J} + e^{-2\alpha\varphi}\right]\left({\tilde \Psi} - \Psi\right)^2=
 4e^{-4{\tilde \nu}-2{\tilde \mu}}\frac{\left({\tilde \Psi} - \Psi\right)^2}{\rho^4} + {\cal O}(\rho^2)
 \end{eqnarray}
and therefore this term is bounded since ${\tilde \Psi} - \Psi={\cal O}(\rho^2)$. The last term in (\ref{SSTARGEN}) is obviously  bounded.

 \medskip
\noindent

(iv) The continuity argument shows that $S_*$ is bounded  in open neighborhoods of the corners.

 \medskip
\noindent

We have shown that $S$ is globally bounded on $\mr$ including the $z$-axis and vanishes at infinity. Therefore $S$ vanishes everywhere and this competes the proof.

 \medskip
\noindent

\section{Conclusion}

In the present paper we addressed the general problem of uniqueness of stationary and asymptotically flat black hole solutions in 5D Einstein-Maxwell-dilaton gravity and in
5D Einstein-Maxwell gravity as a particular case with certain restrictions on the dilaton coupling parameter.  We  have proved that the black hole solutions are fully specified in terms of their interval structure, electric and magnetic charges
as well as the magnetic fluxes. The proofs are based on the non-positivity of the Riemann curvature operator on the space of potentials which in turn insures the non-positiveness of the sectional curvature.
The Riemann curvature operator is however  non-positive not for all values of the dilaton coupling parameter $\alpha$. In other words we imposed restrictions on the dilaton
coupling parameter. The natural question is whether one can escape from these restrictions on $\alpha$ or not. In general sectional curvature can be
non-positive even for positive  Riemann operator. The main difficulty comes from the well known fact that the sections curvature is very poorly studied in differential geometry  and today
we even do not know  where the sectional curvature takes its maxima and minima.  Nevertheless by inspection of the concrete expressions of the sectional curvature
of the potential space considered in the present paper it seems that the sectional curvature is not non-positive everywhere. This fact  makes us to expect
that the black hole solutions, with dilaton coupling parameter outside the restrictions imposed, might exhibit non-uniqueness.  The study of the non-uniqueness
of the black holes  seems to be not less interesting than the study of uniqueness of the black holes and this might open a new  avenue
for investigation (see for example \cite{DYSK}).

The uniqueness theorems of the present work can be easily generalized to the case with multiple disconnected non-degenerate horizons. {\it The new parameters that should be included in the theorems
are the magnetic fluxes between the horizons  through the 2-dimensional surfaces with disk topology or/and through the 2-surfaces with  topology of a disk with a hole in the middle
in addition to the electric charges of the horizons, magnetic charges associated with the finite intervals (different from the horizons) and the left and right magnetic fluxes.}

It seems also possible to extend the uniqueness theorem to the case of rotating asymptotically non-flat Einstein-Maxwell-dilaton (and Einstein-Maxwell) black holes. In particular, the
uniqueness theorem could be extended for rotating Einstein-Maxwell-dilaton black hole solutions with $C$-metric  and Melvin-Ernst asymptotic (see for example \cite{Y3}). In this case
however the orbit space ${\hat M}$ has structure different from that of the asymptotically flat case. More straightforward extension, although subtle in some details, is the extension of our results to the case of extremal black objects which requires an investigation of the near horizon geometry.

The appearance of the magnetic fluxes in the uniqueness theorems makes us  think that the magnetic fluxes should appear also in the thermodynamics of the black
holes and especially  in the expression for the first law as in the case of spacetime with compact dimensions \cite{YN1,YN2}. One also may expect that there are
new exact black hole solutions that  differ from the already know solutions in the magnetic fluxes.

\vskip 1cm
\noindent
{\bf Acknowledgements:}
This work was partially supported by the Bulgarian National Science Fund under Grants DO 02-257, VUF-201/06 and by Sofia University Research Fund.

\end{document}